\documentclass[a4paper]{article}
\usepackage{xcolor}
\usepackage[utf8]{inputenc}
\usepackage[T1]{polski}
\usepackage{graphicx}\graphicspath{{img/}}
\usepackage{subcaption}
\usepackage{lmodern,fancyhdr,secdot,float,amsmath,amsfonts,amsthm,amssymb}
\usepackage{booktabs,tabularx}
\usepackage[margin=2cm]{geometry}
\usepackage{multicol}
\usepackage[hidelinks,unicode]{hyperref}
\usepackage{titlesec}
\titleformat{\section}{\centering\normalsize\bf}{\thesection.}{.5em}{\MakeUppercase}
\titleformat*{\subsection}{\bf\normalsize\selectfont}
\titleformat*{\subsubsection}{\bf\normalsize\selectfont}
\newcommand{\titlePL}[1]{\large\textbf{ #1}}
\newcommand{\titleEN}[1]{\normalsize #1}
\newcommand{\keywordsPL}[1]{\small\textbf{Słowa kluczowe:} #1}
\newcommand{\keywordsEN}[1]{\small\textbf{Keywords:} #1}
\newcommand{\abstractPL}[1]{\small\textbf{Streszczenie:} #1}
\newcommand{\abstractEN}[1]{\small\textbf{Abstract:} #1}

\definecolor{logo_color}{RGB}{40, 69, 166}

\begin{document}\thispagestyle{empty}\pagestyle{fancy}
\begin{minipage}[t]{0.5\textwidth}\vspace{0pt}%
\includegraphics[scale=0.9]{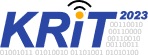}
\end{minipage}
\begin{minipage}[t]{0.45\textwidth}\vspace{12pt}%
\centering
\color{logo_color} KONFERENCJA RADIOKOMUNIKACJI\\ I TELEINFORMATYKI\\ KRiT 2024
\end{minipage}

\vspace{1cm}

\begin{center}
\titlePL{Modelowanie nieliniowej charakterystyki szerokopasmowych wzmacniaczy radiowych o zmiennym napięciu zasilania}

\titleEN{Modeling Nonlinear Characteristics of Wideband Radio Frequency Amplifiers with Variable Supply Voltage}\medskip

Kornelia Kostrzewska$^{1}$;
Paweł Kryszkiewicz$^{2}$

\medskip

\begin{minipage}[t]{0.7\textwidth}
\small $^{1}$ Politechnika Poznańska, Poznań, \href{mailto:email}{kornelia.kostrzewska@student.put.poznan.pl}\\
\small $^{2}$ Politechnika Poznańska, Poznań, \href{mailto:email}{pawel.kryszkiewicz@put.poznan.pl}
\end{minipage}

\medskip

\end{center}

\medskip

\begin{multicols}{2}
\noindent
\abstractPL{
Praca ma na celu zaproponowanie nowego modelu dla nieliniowej charakterystyki wzmacniacza radiowego ze zmiennym napięciem zasilania pracującym w szerokim zakresie częstotliwości. Przedstawiona została propozycja rozszerzonego modelu Rappa. Zaproponowany model zweryfikowano na podstawie pomiarów charakterystyk trzech różnych wzmacniaczy. Model ten może być wykorzystany do projektowania systemów 6G "świadomych" niedoskonałości układów wejściowo-wyjściowych. 
\footnote[1]{Praca powstała w~ramach projektu OPUS  finansowanego przez Narodowe Centrum Nauki nr 2021/41/B/ST7/00136.}}
\medskip

\noindent
\abstractEN{
The work aims to propose a new nonlinear characteristics model for a wideband radio amplifier of variable supply voltage. An extended Rapp model proposal is presented.
The proposed model has been verified by measurements of three different amplifiers. This model can be used to design frontend-aware 6G systems. 
}
\medskip

\noindent
\keywordsPL{Modelowanie wzmacniacza, Model Rappa, Śledzenie obwiedni, Wzmacniacz mocy}
\medskip

\noindent
\keywordsEN{Amplifier modeling, Rapp model, Envelope tracking, Power amplifier}

\section{Wstęp}
Prężny rozwój komunikacji bezprzewodowej jest ściśle związany z odpowiednim dostosowaniem systemów do obsługi coraz większej liczby urządzeń, przy zachowaniu odpowiedniej jakości proponowanych rozwiązań. Jedną z najnowszych odpowiedzi na te rosnące wymagania jest technologia 5G, która przewyższa swoich poprzedników na wielu płaszczyznach, zapewniając nawet kilkadziesiąt razy większą przepustowość komórkową oraz efektywność widmową  \cite{thota_2020_analysis_of_hybrid_PAPR_reduction_methodth_of_OFDM}. Jednym z powodów tak dobrych własności tego systemu jest wykorzystanie złożonych schematów modulacji i technik wielodostępu, np. OFDM (ang. \textit{Orthogonal Frequency Division Multiplexing}) \cite{glock2015memoryless, thota2020analysis}. Zwiększają one przepustowość przy ograniczonym paśmie częstotliwości. Jednak sygnały te charakteryzują się wysokim stosunkiem mocy szczytowej do mocy średniej (ang. \textit{Peak-to-Average Power Ratio}, PAPR) i stawiają duże wymagania co do liniowości układu nadajnika w celu utrzymania zniekształceń sygnału na niskim poziomie.

Kluczowym, w tej kwestii, elementem w torze nadawczym jest wzmacniacz mocy (ang. \textit{Power Amplifier}, PA). 
Ich wykorzystanie jest związane ze znacznym zużyciem energii, a wprowadzane przez niego zniekształcenia nielinowe mogą przyczyniać się do zniekształceń sygnału w całym pasmie. Konieczna jest zatem wiedza o podstawowych parametrach PA np. jego liniowości. Celem jest zidentyfikowanie najkorzystniejszego punktu pracy będącego kompromisem pomiędzy efektywnością energetyczną, a efektywnością widmową.

W tym celu wykorzystuje się różne podejścia do architektury układów wzmacniaczy \cite{joung2014survey}, które mają na celu osiągnięcie jak najbardziej liniowego wzmocnienia sygnału przy jednoczesnej minimalizacji zużycia energii. Śledzenie obwiedni (ang. \textit{Envelope Tracking}, ET) wyróżnia się jako jedno z najbardziej popularnych i bardzo skutecznych podejść \cite{kim2011optimization_for_ET_shaped_operation_od_ETPA}. Podstawowym założeniem jest zmienianie amplitudy napięcia zasilającego wzmacniacz, do chwilowej obwiedni sygnału podawanego na tor radiowy wzmacniacza. 
 
Z perspektywy projektowania algorytmów nadawczo-odbiorczych lub optymalizacji parametrów transmisyjnych konieczna jest znajomość charakterystyki nieliniowej wzmacniacza. Najczęściej odbywa się to poprzez modelowanie behawioralne \cite{kim2011optimization_for_ET_shaped_operation_od_ETPA}. Modele behawioralne nie wymagają głębokiej wiedzy na temat budowy wewnętrznej wzmacniacza np. układu tranzystorów. Skupiają się one na sygnale podawanym na wejście i uzyskanym na wyjściu wzmacniacza w wyniku pomiarów lub symulacji obwodów elektronicznych \cite{Ghannouchi_behavioral_modeling_and_predistortion_2009, glock2015memoryless}.

Istnieje wiele publikacji naukowych \cite{Ghannouchi_behavioral_modeling_and_predistortion_2009, joung2014survey, kryszkiewicz_2018_amplifier-coupled_tone_reservation, thota_2020_analysis_of_hybrid_PAPR_reduction_methodth_of_OFDM}, które szeroko omawiają podstawowe modele behawioralne. Niektóre modelują tylko efekty bezpamięciowe (Soft limiter, Rapp, Saleh). Inne uwzględniają również „pamięć” wzmacniacza (Voltera, Wiener, Hammerstein). Wybór w dużej mierze zależy od badanego w danym przypadku wzmacniacza oraz wymaganej dokładności.

Jako że technika ET upowszechnia się pozwalając osiągać wysoką jakość nadawanego sygnału przy wysokiej efektywności energetycznej konieczne jest opracowanie nowych modeli behawioralnych uwzględniających zmienne napięcie zasilania jak również szerokopasmowy charakter transmisji oczekiwany od systemów 6G. W pracach \cite{al-kanan_2017_extended_saleh_model_for_behavioral_modeling_ET_PA, al2018hysteresis_nonliearity_model} zaprezentowane są propozycje rozszerzonego modelu Saleha. Autorzy dodają niezależne współczynniki, które poprzez funkcję wielomianową uzależniają charakterystykę nieliniową od chwilowego napięcia zasilania. 

Z drugiej strony, w \cite{li2016new_model_for_ET_PA, mengozzi2021joint_dual_input_DP} punktem wyjścia jest tradycyjny pamięciowy model wielomianowy, w którym współczynniki są uzależniane od obwiedni sygnału na wejściu wzmacniacza i wejścia modulatora zasilania. Przeanalizowano także prace badawcze koncentrujące się na rozszerzeniu modelu Rappa \cite{tafuri_2015_memory_models_for_behavioral_modeling_and_DPD_of_ET_PA, wang2005design_of_Widebans_ET_PA_for_OFDM_application}. Podobnie jak poprzednio napięcie zasilania jest włączone do modelu podstawowego jako dodatkowa zmienna, która wpływa na współczynniki definiujące nieliniowość wzmacniacza.

Wszystkie powyższe prace skupiają się na włączeniu do modelu napięcia zasilania PA. Jednak, z uwagi na to, że współczesne systemy telekomunikacyjne działają w szerokim zakresie częstotliwości, włączenie częstotliwości nośnej jako parametru do analizy sygnału może korzystnie wpłynąć na efektywność widmową, a także energetyczną systemu. W żadnym z wymieniowych powyżej artykułów nie rozważano tego problemu.

Ta praca przedstawia nowy model behawioralny oparty na tradycyjnym modelu Rappa, który został rozszerzony o wpływ zmiennego napięcia zasilania wzmacniacza, a także częstotliwości nośnej sygnału nadawanego. Model Rappa został wybrany z uwagi na swoją prostotę, wyrażoną niewielką liczbą parametrów, oraz dużą popularność, wynikająca z dobrego dopasowania do współczesnych wzmacniaczy mocy wykorzystujących tranzystory, \cite{Glock2015}. Został też zaproponowany do analizy wpływu wzmacniacza na sygnał systemu 5G \cite{Nokia_3gpp_Rapp}.  

W rozdziale \ref{sec:system_pomiarowy} zaprezentowany został wykorzystany system pomiarowy oraz wyniki pomiarów. W rozdziale \ref{sec:proponowany model} przedstawiony jest proponowany model charakterystyki wzmacniacza. Analiza poprawności zaproponowanego modelu zamieszczona została w rozdziale \ref{sec:wyniki}, a wnioski znajdują się w rozdziale \ref{sec:conclusions}.

\section{System pomiarowy i wyniki pomiarów} \label{sec:system_pomiarowy}

\begin{figure}[H]
\centering
\includegraphics[scale=0.35]{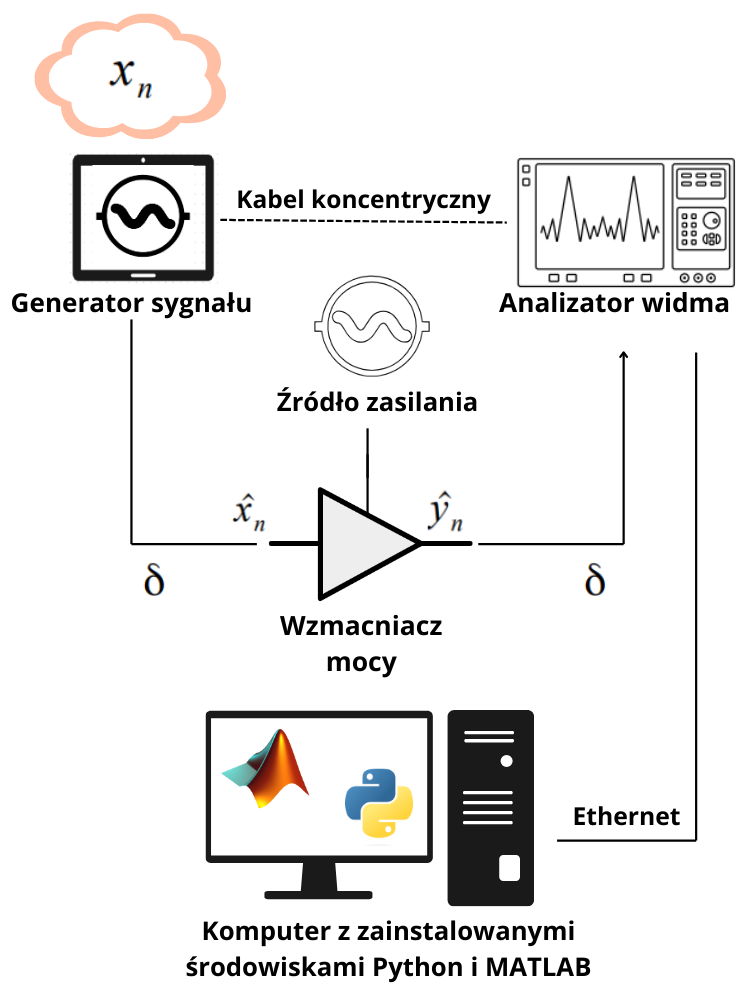}
\caption{Schemat pomiarowy}
\label{fig:scheme-set-up}
\end{figure}

Schemat pomiarowy przedstawiono na Rys. \ref{fig:scheme-set-up} i składa się z następujących elementów:
\begin{itemize}
\item generator sygnału: Rohde \& Schwarz SMBV 100A,
\item analizator widma: Rohde \& Schwarz FSL 6 \cite{FSL6_manual},
\item badane wzmacniacze mocy od Mini-Circuits: ZFL-2000+ \cite{nota_ZFL-2000+}, ZX60-5916 \cite{nota_ZX60-5916}, ZX60-2534 \cite{nota_ZX60-2534},
\item źródło zasilania: NN M10-QP-305E,
\item komputer z zainstalowanymi środowiskami MATLAB i Python, połączony za pomocą kabla Ethernetowego.
\end{itemize}
Głównym celem tego zestawu jest zebranie próbek po ich zniekształceniu przez wzmacniacz mocy. 

Generator sygnału wysyła cyklicznie 10 symboli OFDM, o zadanej częstotliwości, które są podawane na wejście testowanego wzmacniacza. Jeden symbol składa się z 4096 podnośnych, spośród których zajętych jest 590 podnośnych z zakresu od -300 do -1 oraz od 10 do 300.

Wzmacniacz zasilany jest napięciem stałym z zasilacza. Dostępny układ pomiarowy nie pozwalał na dynamiczną zmianę napięcia zasilania jak w przypadku architektury ET. 
Napięcie zasilania zmieniane jest z pewnym krokiem zależnym od modelu wzmacniacza (co opisano w rozdziale  \ref{sec:zakres_parametro}) w całym dostępnym dla danego modelu przedziale ręcznie. Podobnie zmieniano częstotliwość nośną nadawanego sygnału. Zabieg ten ma na celu uzyskanie pełnej charakterystyki zachowania wzmacniacza.

Sygnał z wyjścia wzmacniacza jest odbierany przez analizator widma. Pobrane próbki mają formę zespolonych próbek IQ, które umożliwiają analizę  sygnału w paśmie podstawowym. Podejście to jest powszechne w wielu badaniach dotyczących analizy lub modelowania sygnałów  \cite{boumaiza_2007_systematic_and_adaptive_characterization_for_behavioral_modeling_of_dynamic_nonlinear, kryszkiewicz_2015_obtaining_low_out_band_emission_of_NC-OFDM, kryszkiewicz_2018_amplifier-coupled_tone_reservation}. W odebranym wektorze próbek IQ zapisany jest jeden pełny okres sygnału nadawanego. Aby uniknąć zniekształceń nieliniowych na wejściu analizatora widma, konieczne było ustawienie silnego tłumienia na wejściu analizatora, tzn. około $50 dB$.

Warto również wspomnieć, że oscylatory lokalne nadajnika i odbiornika zostały zsynchronizowane poprzez kabel koncentryczny. 

Ostatecznie odebrany wektor próbek IQ trafia do komputera podłączanego do analizatora za pomocą kabla Ethernetowego.

Przygotowanie próbek obejmowało uwzględnienie tłumienia kabli łączących urządzenia oraz tzw. 'wirtualnego wzmocnienia' w przypadku sygnału wejściowego, wynikającego z faktu, że próbki cyfrowe sygnału nadawanego są „wzmacniane” poprzez generator sygnałowy do określonej mocy. Dokładny sposób pomiaru tłumienia oraz przeskalowanie próbek sygnału wejściowego i wyjściowego przedstawiono w \cite{kostrzewska2024power}.

W następnym kroku próbki poddano korekcji przesunięcia w dziedzinie czasu oraz w dziedzinie częstotliwości. Wykorzystana procedura korekcji została przedstawiona w \cite{kostrzewska2024power}. 
W efekcie dysponujemy dwoma sygnałami: wejściowym $\hat{x}$ oraz wyjściowym ${y_n}$ które są zależne poprzez funkcję $g(~)$ stanowiąca charakterystykę wzmacniacza: 
\begin{equation}
\label{output_of_PA_after_corr}
{y_n}=g\left(\hat{x}\right).
\end{equation}
Możliwe jest więc dopasowanie modelu behawioralnego, jak pokazano w \cite{kostrzewska2024power}.

Dla każdego zestawu parametrów wejściowych dopasowano do próbek wyjściowych model Rappa. W tym modelu sygnał (zespolony) na wyjściu wzmacniacza jest opisywany jako:
\begin{equation}
\label{Rapp_model_formula_base}
    {y_n} = G{\hat{x_n}} \left ( 1+ \left ( \frac{\left |{\hat{x_n}}\right |}{V_{sat}}\right) ^{2 p}\right) ^{-\frac{1}{2p}},
\end{equation}
gdzie $p$ to współczynnik gładkości (liczba rzeczywista dodatnia), $V_{sat}$ napięcie nasycenia na wejściu wzmacniacza, $G$ to wzmocnienie liniowe amplitudy sygnału, a $\hat{x_n}$ odnosi się do sygnału podawanego na wejście wzmacniacza. 

\begin{figure}[H]
\centering
\includegraphics[width=3.4in]{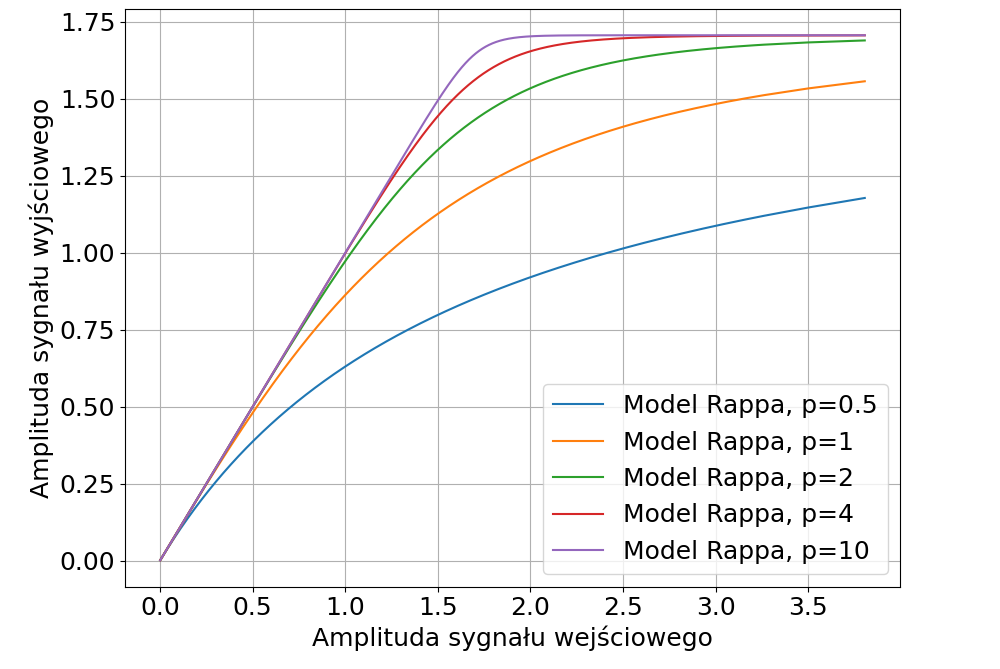}
\caption{Charakterystyka AM/AM dla różnych wartości $p$, $V_{sat} = 1.7 V$, $G = 1$ w skali liniowej 
(V). }
\label{fig:Rappp_klka_p}
\end{figure}

Rys. \ref{fig:Rappp_klka_p} przedstawia charakterystykę AM/AM wzmacniacza tzn. amplitudę sygnału wyjściowego w funkcji amplitudy sygnału wejściowego, modelowanego wzorem (\ref{Rapp_model_formula_base}) dla różnych wartości $p$. Napięcie nasycenia $V_{sat}$ oraz wzmocnienie $G$ zostały przyjęte jako stałe. Zauważalne jest, że im większa wartość parametru $p$, tym w większym zakresie amplitud wzmacniacz działa liniowo, z wyraźnym obcięciem wyższych amplitud. 

\subsection{Zakresy parametrów wejściowych wzmacniaczy}
\label{sec:zakres_parametro}

Dla każdego modelu badanego wzmacniacza, na podstawie danych przedstawionych w jego nocie katalogowej, dobrano zakres napięcia zasilającego wzmacniacz (Tabela \ref{pa_supply_range_in_power}) oraz zakres częstotliwości nośnej (Tabela \ref{pa_freq_range_in_power}). 

\begin{table}[H]
\centering
\caption{Zakres napięcia zasilania i testowane wartości dla wzmacniaczy mocy}
\label{pa_supply_range_in_power}
\medskip
\begin{tabularx}{\columnwidth}{llX}
\toprule
\textit{Wzmacniacz} & \textit{Katalogowe} & \textit{Testowane }\\ 
\textit{} & \textit{wartości} & \textit{wartości}\\
\textit{} & \textit{ [V]} & \textit{[V]}\\\midrule
    ZFL-2000+ & max 17 & 8.0,9.5,…, 14.5,15.0  \\
    ZX60-5916 & 2.8, 5.0& 2.4,2.6,…, 4.8,5.0\\
    ZX60-2534 & 2.8, 5.0& 2.4,2.6,…, 4.8,5.0\\ \bottomrule
\end{tabularx}
\end{table}

\begin{table}[H]
\centering
\caption{Zakres częstotliwości nośnej i testowane wartości dla wzmacniaczy}
\label{pa_freq_range_in_power}
\medskip
\begin{tabularx}{\columnwidth}{llX}
\toprule
\textit{Wzmacniacz} & \textit{Katalogowe} & \textit{Testowane }\\ 
\textit{} & \textit{wartości } & \textit{wartości}\\
\textit{} & \textit{ [MHz]} & \textit{[GHz]}\\\midrule
    ZFL-2000+ & 10 – 2000  &	 0.5,1,…,2   \\
    ZX60-5916 & 1500 — 6000& 2,2.5,…,5.5 \\
    ZX60-2534 & 500 — 2500& 0.5,1,…,2.5 \\ \bottomrule
\end{tabularx}
\end{table}

\subsection{Analiza wyników pomiarów}
Pomiary oraz korekcje przeprowadzono dla każdego modelu w całym zakresie jego parametrów. Następnie wyznaczono współczynniki modelu Rappa, wykorzystując do tego algorytm przedstawiony w \cite{kostrzewska2024power}. Wszystkie otrzymane wyniki zostały udostępnione w \cite{kryszkiewicz_2024_10519547}. 

\begin{figure}[H]
\centering
\begin{minipage}{0.3\textwidth}
\begin{subfigure}{\textwidth}
    \includegraphics[scale=0.25]{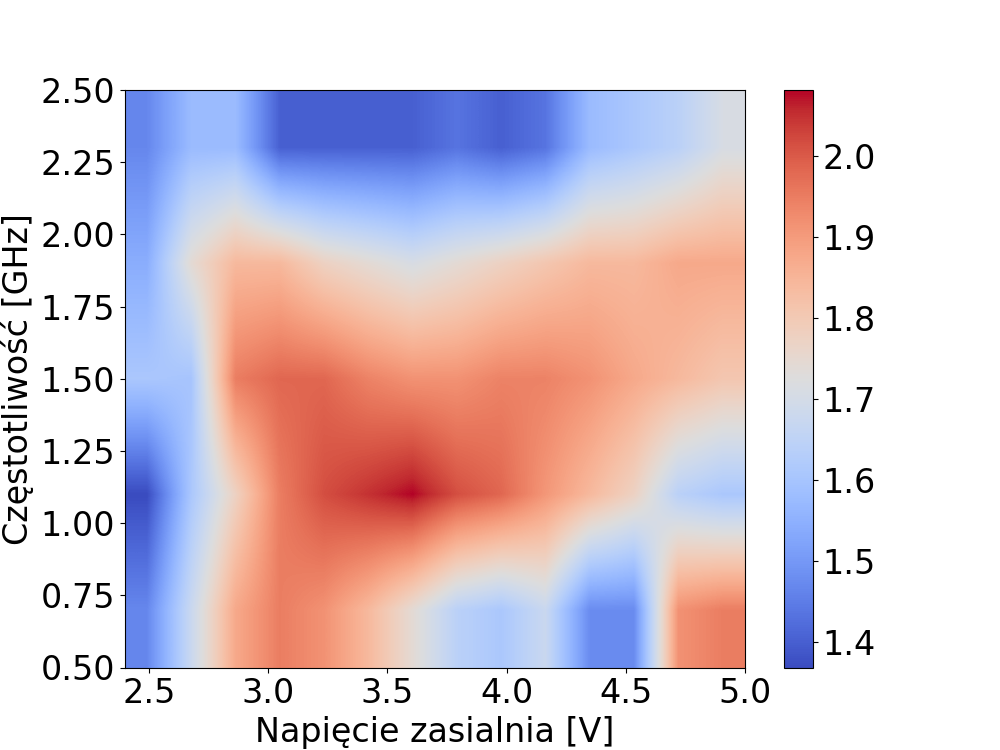}
    \caption{ Parametr $p$}
    \label{fig:heatmapa_p_2000+}
\end{subfigure}
\hfill
\begin{subfigure}{\textwidth}
    \includegraphics[scale=0.25]{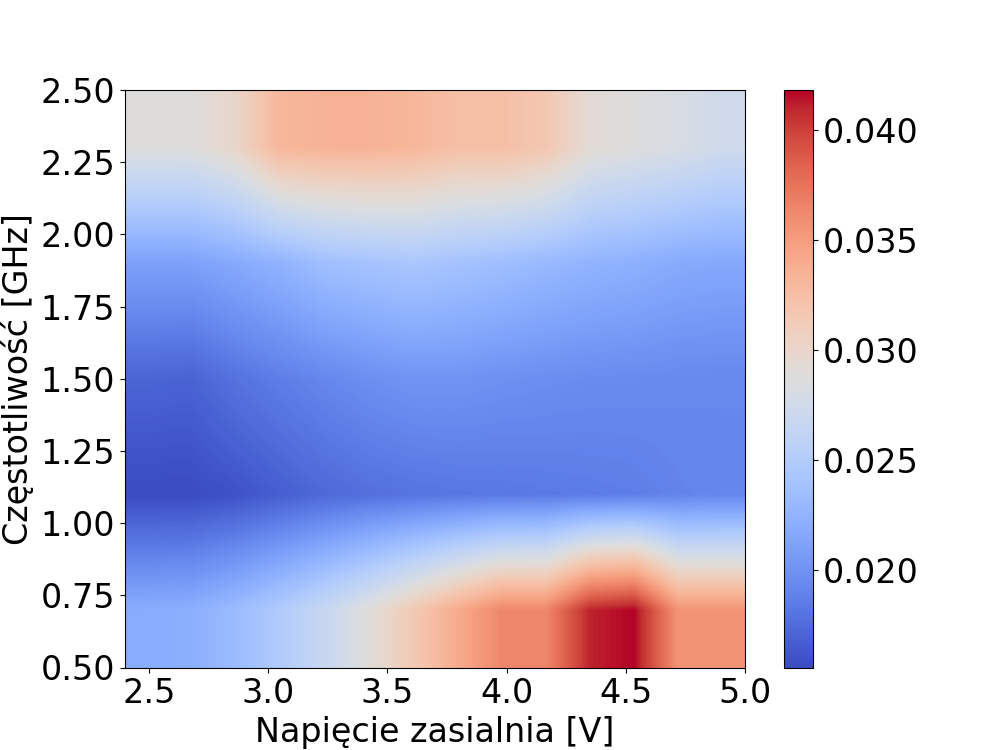}
    \caption{Parametr $V_{sat}$}
    \label{fig:heatmapa_p_2534}
\end{subfigure}
\hfill
\begin{subfigure}{\textwidth}
    \includegraphics[scale=0.25]{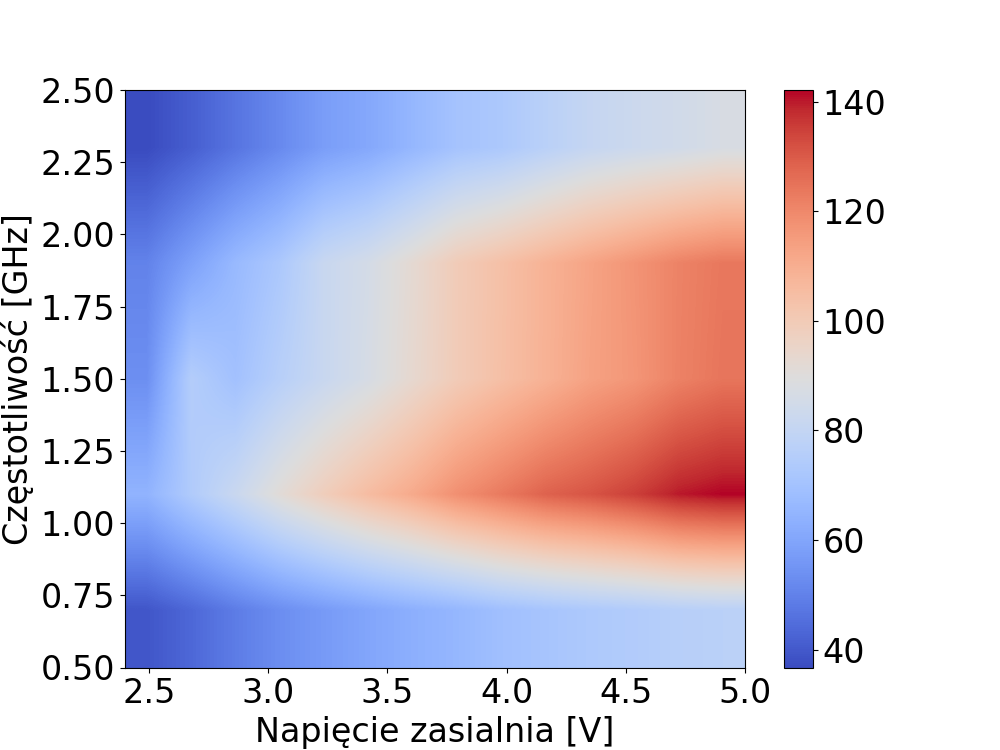}
    \caption{Parametr $G$}
    \label{fig:heatmapa_p_5916}
\end{subfigure}
\end{minipage}
\caption{Mapy cieplne dla modelu ZX60-2534}
\label{fig:heatmaps_p}
\end{figure}

Rys. \ref{fig:heatmaps_p} przedstawia otrzymane wartości współczynników Rappa dla wzmacniacza ZX60-2534 oraz ich zmienność w funkcji parametrów wejściowych za pomocą map cieplnych. 

Widoczne jest, że wzmocnienie rośnie wraz z napięciem zasilania. Tempo tego wzrostu jest jednak istotnie różne w zależności od częstotliwości nośnej. Należy pamiętać, że jest to wzmocnienie napięcia w skali liniowej. Analizując zmienność współczynnika gładkości $p$ zauważalna jest znacząca zmienność wartości tego parametru. Pomimo istotnych różnic między wzmacniaczami (wyniki dla pozostałych dwóch wzmacniaczy mogą być znalezione w \cite{kostrzewska2024power}), maksymalne wartości $p$ dla wszystkich przypadków nie przekraczają $p=3,0$. Oznacza to, że badane wzmacniacze wykazują silnie nieliniowe zachowanie. Natomiast napięcie nasycenia $V_{sat}$ zmienia się wraz z parametrami wejściowymi co najmniej dwukrotnie dla każdego badanego modelu wzmacniacza. Ogólnie można powiedzieć, że jego zmienność stanowi funkcję dwóch badanych parametrów wejściowych. Podsumowując, każdy z obserwowanych parametrów zmienia się znacząco zarówno w funkcji częstotliwości jak i napięcia zasilania wzmacniacza co oznacza konieczność ich zamodelowania. Wykresy nie wykazują jednak gwałtownych zmian (nieciągłości), co jest podstawą do zaproponowanego rozszerzenia modelu Rappa, w którym wartości współczynników wyrażone są przez funkcje dwóch zmiennych wejściowych, co pokazano w rozdziale \ref{sec:proponowany model}.

%

\section{Proponowany model charakterystyki wzmacniacza}
\label{sec:proponowany model}

Przedstawiony w tej pracy model został zainspirowany rozszerzeniem modelu Rappa, przedstawionym w \cite{tafuri_2015_memory_models_for_behavioral_modeling_and_DPD_of_ET_PA}. Jego autorzy opisali współczynniki modelu $G$, $p$ oraz $V_{sat}$  jako funkcje zależne od obwiedni napięcia zasilania PA $V_{sup}$:
\begin{align}
\label{Rapp_model_formula_ext1}
    y_n\!=\! G(V_{sup})\hat{x}_n 
    \left (\! 1\!+ \! \left (\! \frac{\left |\hat{x}_n\right|}{V_{sat}(V_{sup})}\!\right)^{2 p(V_{sup})}\right)^{-\frac{1}{2p(V_{sup})}}
\end{align}

Nasza propozycja to uwzględnienie dodatkowo wartości częstotliwości nośnej w funkcjach opisujących wartości współczynników modelu Rappa tzn. 
\begin{equation}
\label{Rapp_model_formula_ext1}
\centering
    y_n\!\!=\!G(V_{sup},f)\hat{x_n} \!\!
    \left ( \!\!1\!\!+\!\! \left (\! \frac{\left |\hat{x_n}\right |}{V_{sat}(V_{sup},f)}\!\right)^{\!\!2 p(V_{sup},f)}\!\right)^{\!\!-\frac{1}{2p(V_{sup},f)}}
\end{equation}

Pierwszym krokiem było przeprowadzenie pomiarów 3 różnych wzmacniaczy dla określonych wartości napięcia zasilania wzmacniacza $V_{sup}$ oraz częstotliwości nośnej $f$ i wyznaczenie dla każdej badanej pary wartości współczynników modelu Rappa: $p$, $V_{sat}$ oraz $G$. Następnie na tej podstawie dopasowano funkcje opisujące te parametry. W tym celu wykorzystanie jedno z rozszerzeń środowiska MATLAB - MATLAB Curve Fitting Tool.

W przypadku każdego współczynnika przetestowano kilka funkcji szukając takiej o najlepszym dopasowaniu dla wzmacniacza ZX60-2534. 

Pierwsze podejście polegało na dopasowaniu powierzchni zdefiniowanej przez wielomian drugiego stopnia zarówno w funkcji $f$ jak i $V_{sup}$, niezależnie dla współczynników $p$, $V_{sat}$ oraz $G$,
\begin{equation}
\begin{aligned}
poly22 &= p00 + p10 \cdot V_{sup} + p01 \cdot f + p20 \cdot V_{sup}^2 +\\& + p11 \cdot V_{sup} \cdot f + p02 \cdot f^2.
\end{aligned}
\label{poly22}
\end{equation}
Rozważono także wielomian trzeciego stopnia:
\begin{equation}
\begin{aligned}
poly33 &= p00 + p10 \cdot V_{sup} + p01 \cdot f + p20 \cdot V_{sup}^2  +\\ & + p11 \cdot V_{sup} \cdot f  + p02 \cdot f^2 + p30 \cdot V_{sup}^3+ \\ &+ p21 \cdot V_{sup}^2 \cdot f + p12 \cdot V_{sup} \cdot f^2 + p03 \cdot f^3.
\end{aligned}
\label{poly33}
\end{equation}

Następne podejście polegało na dopasowaniu dwóch niezależnych wielomianów, uwzględniających zależność od $V_{sup}$ oraz $f$, i wyrażanie całkowitej zależności jako iloczyn równań opisujących te charakterystyki tzn. 
\begin{equation}
f(V_{sup}, f) = g(V_{sup}) \cdot h(f).
\label{vrazyf}
\end{equation}

Dla parametrów $V_{sat}$ oraz $G$ funkcje te mają więc postaci:
\begin{equation}
f_{V_{sat}}(V_sup, f) = (log(V_{sup})+a) \cdot (b \cdot f^3+c \cdot f+d),
\label{polyv4}
\end{equation}
oraz
\begin{equation}
f_{G}(V_sup, f) = (log(V_{sup})+a) \cdot (b \cdot f^3+c \cdot f^2+d \cdot f+e).
\label{polyv4}
\end{equation}
W przypadku parametru $p$ nie udało się dobrać optymalnych funkcji opisujących zmienność oddzielnie dla parametrów $V$ i $f$. Złożoność zmienności tego parametru jest na tyle wysoka, że nie można jej precyzyjnie opisać oddzielnie dla zmiennych wejściowych, ponieważ wykazuje ona silną zależność zarówno od napięcia, jak i częstotliwości.


W ostatnim podejściu skupiono się na stopniowym eliminowaniu kolejnych współczynników wielomianu dla optymalizacji dopasowania. Proces ten rozpoczął się od zdefiniowania funkcji złożonej ze wszystkich możliwych monomianów do stopnia $N$ składających się z kombinacji zmiennych $V_{sup}$ oraz $f$. $N$ dobrano tak wysokie, żeby wartość 
pierwiastka z błędu średniokwadratowego (ang. \textit{Root Mean Square Error}, RMSE) osiągnęła minimum tzn. dalsze zwiększanie stopnia wielomianu nie wpłynęło na obniżenie RMSE.
Następnie współczynniki były usuwane jeden po drugim, obserwując, jak zmieniały się wartości RMSE. Najmniej istotny człon wielomianu tzn. ten, którego usunięcie najmniej podniosło RMSE, był trwale usuwany z funkcji. Te operacje były iteracyjnie powtarzane, dopóki wartości RMSE pozostawały na podobnym poziomie. 

\begin{figure}[H]
\centering
\includegraphics[scale=0.33]{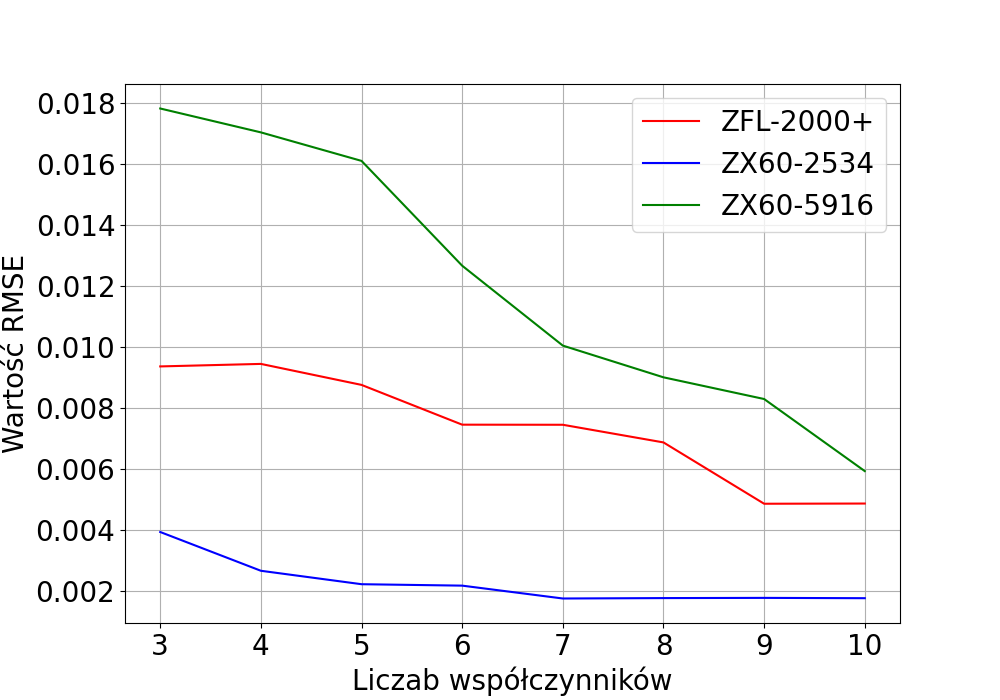}
\caption{RMSE jako funkcja liczby współczynników dla parametru $V_{sat}$}
\label{fig:RMSE_V}
\end{figure}

Przykładowe wyprowadzenie wzoru wykorzystując ostatnie podejście przedstawiono dla współczynnika $V_{sat}$ (analogicznie wyglądało ono dla pozostałych współczynników). W tym przypadku danymi, na których dokonywano redukcji współczynników były te otrzymane z pomiaru wzmacniacza ZX60-2534, a wielomianem początkowym był (\ref{poly33}) o 10 współczynnikach. Na Rys. \ref{fig:RMSE_V} przedstawiono wartości RMSE dla otrzymanych równań. Równania te następnie zaaplikowano do danych z pozostałych dwóch wzmacniaczy. Otrzymane wartości RMSE również przedstawiono na Rys.\ref{fig:RMSE_V}. Można zauważyć, że we wszystkich przypadkach wartość RMSE zwiększa się w miarę usuwania kolejnych wyrazów wielomianu, a dodatkowo jest różna pomiędzy wzmacniaczami. Celem było więc wybranie wielomianu o małej liczbie współczynników, który dobrze dopasowuje się dla wszystkich rozważanych wzmacniaczy. W tym przypadku wybór padł na wielomian opisany czterema współczynnikami, 
\begin{equation}
poly_{V4} = p10 \cdot V_{sup}+p11 \cdot V_{sup} \cdot f+p02 \cdot f^2+p12 \cdot V_{sup} \cdot f^2.
\label{polyv4}
\end{equation}


Aby uzyskać wzór najlepiej opisujący zmienność danego parametru modelu Rappa, porównano równania uzyskane z różnych podejść, uwzględniając wartości RMSE. Postać tych funkcji (np. zbiór monominanów) wyznaczono na podstawie danych ze wzmacniacza dla którego zmienność danego parametru (np. $p$) była największa. Następnie, równania te zastosowano do pozostałych dwóch wzmacniaczy, co umożliwiło porównanie dokładności dopasowania między nimi.

\begin{figure}[H]
\centering
\includegraphics[scale=0.33]{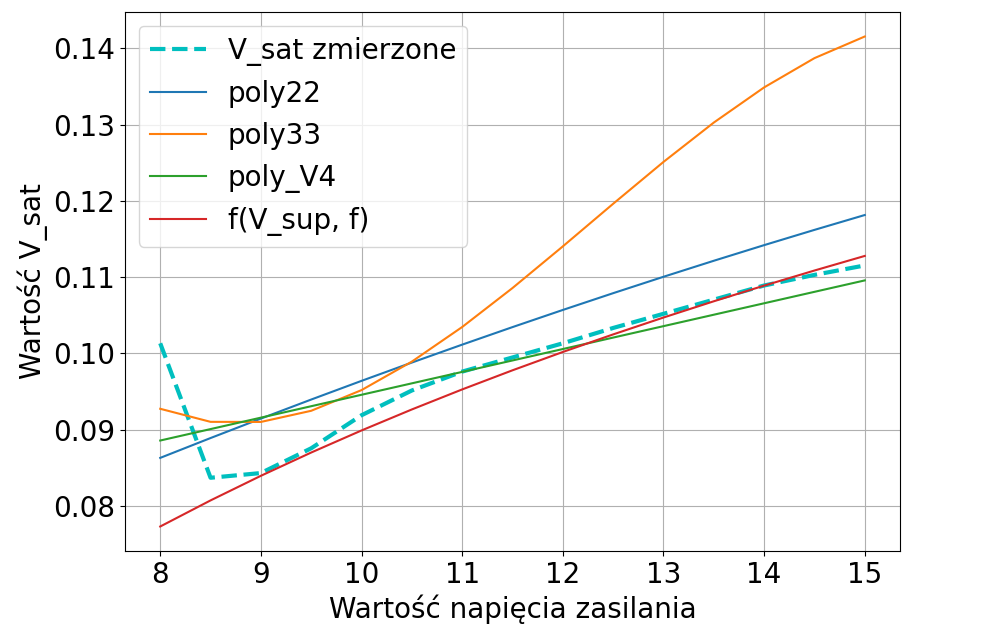}
\caption{Porównanie modelowania $V_{sat}$ dla częstotliwości 1.5GHz}
\label{fig:V-comparison}
\end{figure}

Przykładowe porównanie modelowania współczynnika $V_{sat}$ przedstawiono w Tabeli \ref{v_sat_modelling}, oraz na Rys. \ref{fig:V-comparison} dla wzmacniacza ZFL-2000+ transmitującego sygnał na nośnej 1.5~GHz. Widoczne jest, że wartości RMSE dla wzmacniacza ZX60-2534 są najniższe. Wynika to z faktu, że dane pomiarowe z tego wzmacniacza zostały wykorzystane do wyznaczania postacji funkcji opisującej parametr $V_{sat}$. Funkcje $poly22$ (\ref{poly22}) oraz $f(V_{sup}, f)$ (\ref{vrazyf}) dla każdego wzmacniacza osiągają największe wartości RMSE, natomiast najlepsze dopasowanie wykazuje równanie $poly33$ (\ref{poly33}), ale kosztem wykorzystania największej liczby współczynników. Równanie $poly_{V4}$ (\ref{polyv4}) wykorzystuje jedynie cztery współczynniki i osiąga mniejsze wartości RMSE, niż równanie $f(V_{sup}, f)$ (\ref{vrazyf}), które wykorzystuje tą samą liczbę współczynników, dla dwóch z trzech badanych wzmacniaczy. Widać również, że krzywa wyznaczona przez równanie $poly_{V4}$ (\ref{polyv4}) dobrze pokrywa się z danymi pomiarowymi, dlatego zostało ono wybrane jako najlepiej opisujące zmienność parametru:
\begin{equation}
    V_{sat}(V_{sup}, f) = p10 \cdot V_{sup}+p11 \cdot V_{sup} \cdot f+p02 \cdot f^2+p12 \cdot V_{sup} \cdot f^2.
    \label{poly_V}
\end{equation}

Analogicznie przeprowadzono modelowanie parametrów $p$ oraz $G$. Uzyskano następujące funkcje:
\begin{equation}
    G(V_{sup}, f) = (log(V_{sup})+a) \cdot (b \cdot f^3+c \cdot f^2+d \cdot f+e),
\label{eqG}
\end{equation}
\begin{equation}
    p(V_{sup}, f) = p10 \cdot V_{sup}+p01 \cdot f+p20 \cdot V_{sup}^2+p02 \cdot f^2.
    \label{poly_p2}
\end{equation}

\begin{table}[H]
\centering
\caption{Modelowanie $V_{sat}$}
\label{v_sat_modelling}
\medskip
\begin{tabularx}{\columnwidth}{XXXXX}
\toprule
\textit{Wzmacniacz} & \textit{Równanie} & \textit{Liczba wsp.}& \textit{RMSE}\\\midrule
    ZFL-& $poly22$ & 6  & 0.007\\
    2000+&  $poly33$ & 10  & 0.004\\
    & $poly_{V4}$ & 4  & 0.009\\
    & $f(V_{sup}, f)$ & 4  & 0.014\\\midrule

    ZX60- & $poly22$ & 6  & 0.002 \\
    2534& $poly33$ & 10  & 0.001 \\
    & $poly_{V4}$ & 4  & 0.002  \\
    & $f(V_{sup}, f)$ & 4  & 0.003  \\\midrule

    ZX60- & $poly22$ & 6  & 0.011 \\
    5916& $poly33$ & 10  & 0.005\\
    & $poly_{V4}$ & 4  & 0.017  \\
    & $f(V_{sup}, f)$ & 4  & 0.014  \\  \bottomrule
\end{tabularx}
\end{table}

\section{Analiza jakości dopasowania proponowanego modelu}
\label{sec:wyniki}

\begin{figure}[H]
\centering
\begin{minipage}{0.5\textwidth}
\begin{subfigure}{\textwidth}
    \includegraphics[scale=0.25]{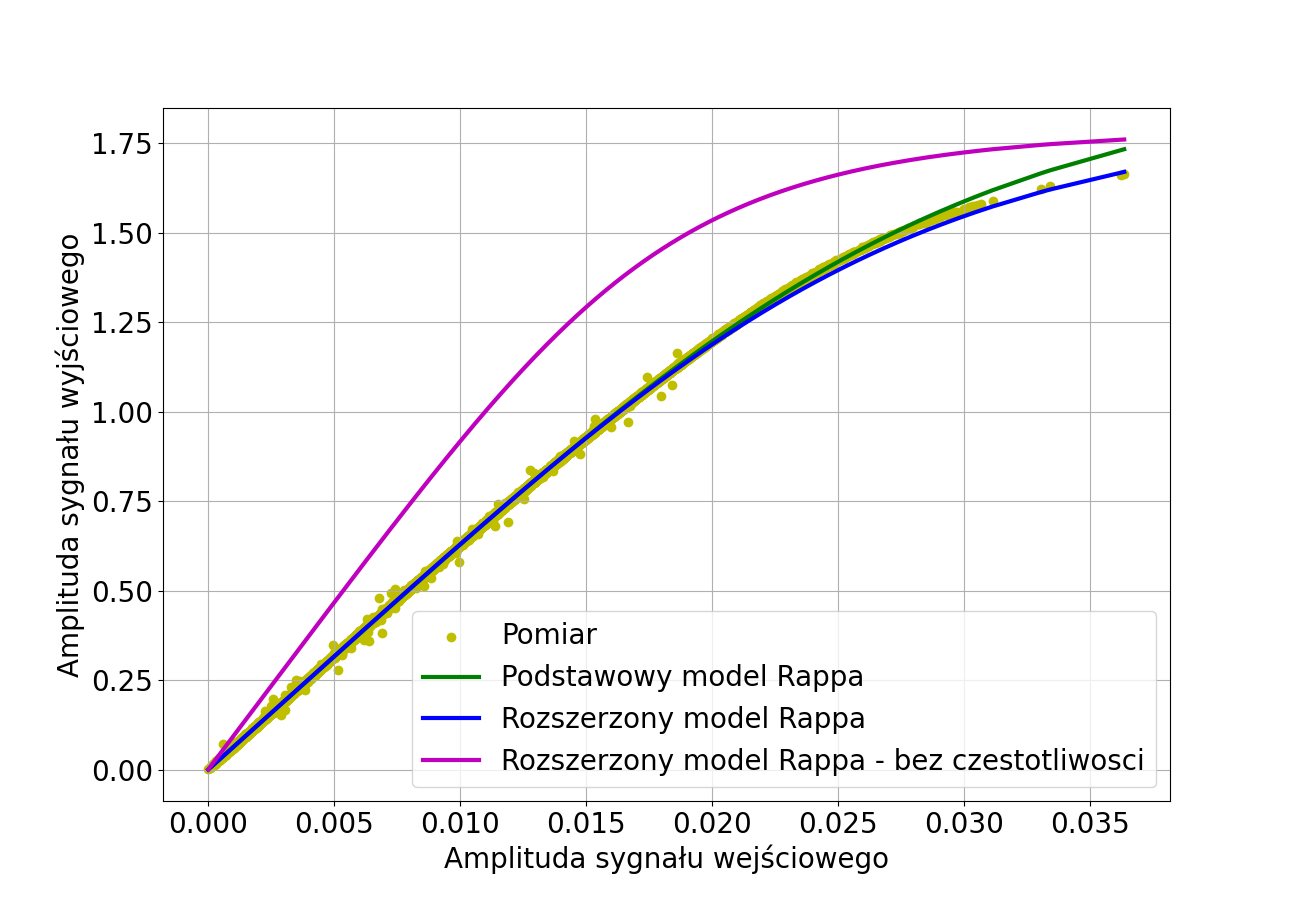}
    \caption{Częstotliwość $f = 0.5GHz$}
    \label{fig:model_20001}
\end{subfigure}
\hfill
\begin{subfigure}{\textwidth}
    \includegraphics[scale=0.25]{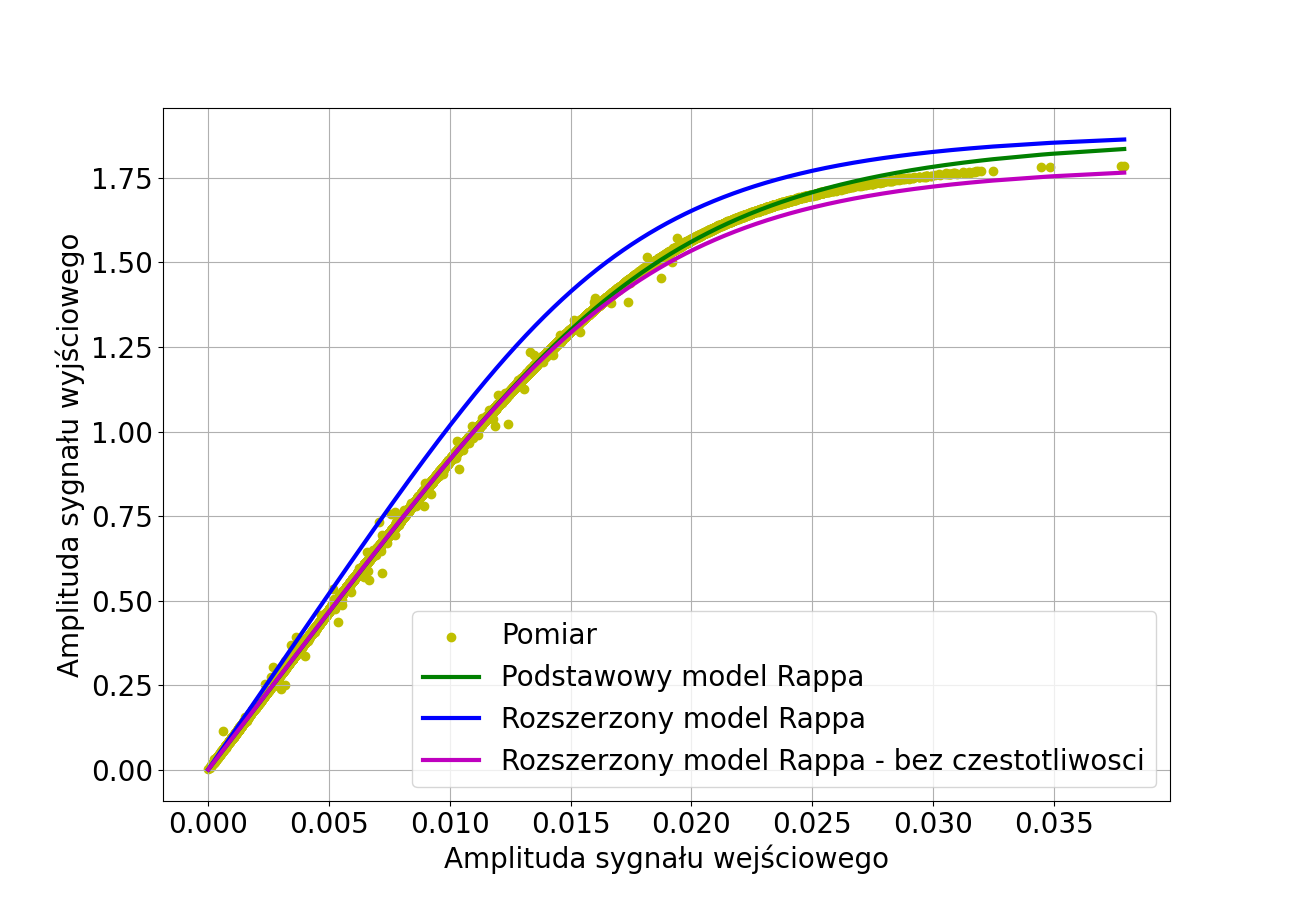}
    \caption{Częstotliwość $f = 1.5GHz$}
    \label{fig:model_25342}
\end{subfigure}
\hfill
\begin{subfigure}{\textwidth}
    \includegraphics[scale=0.25]{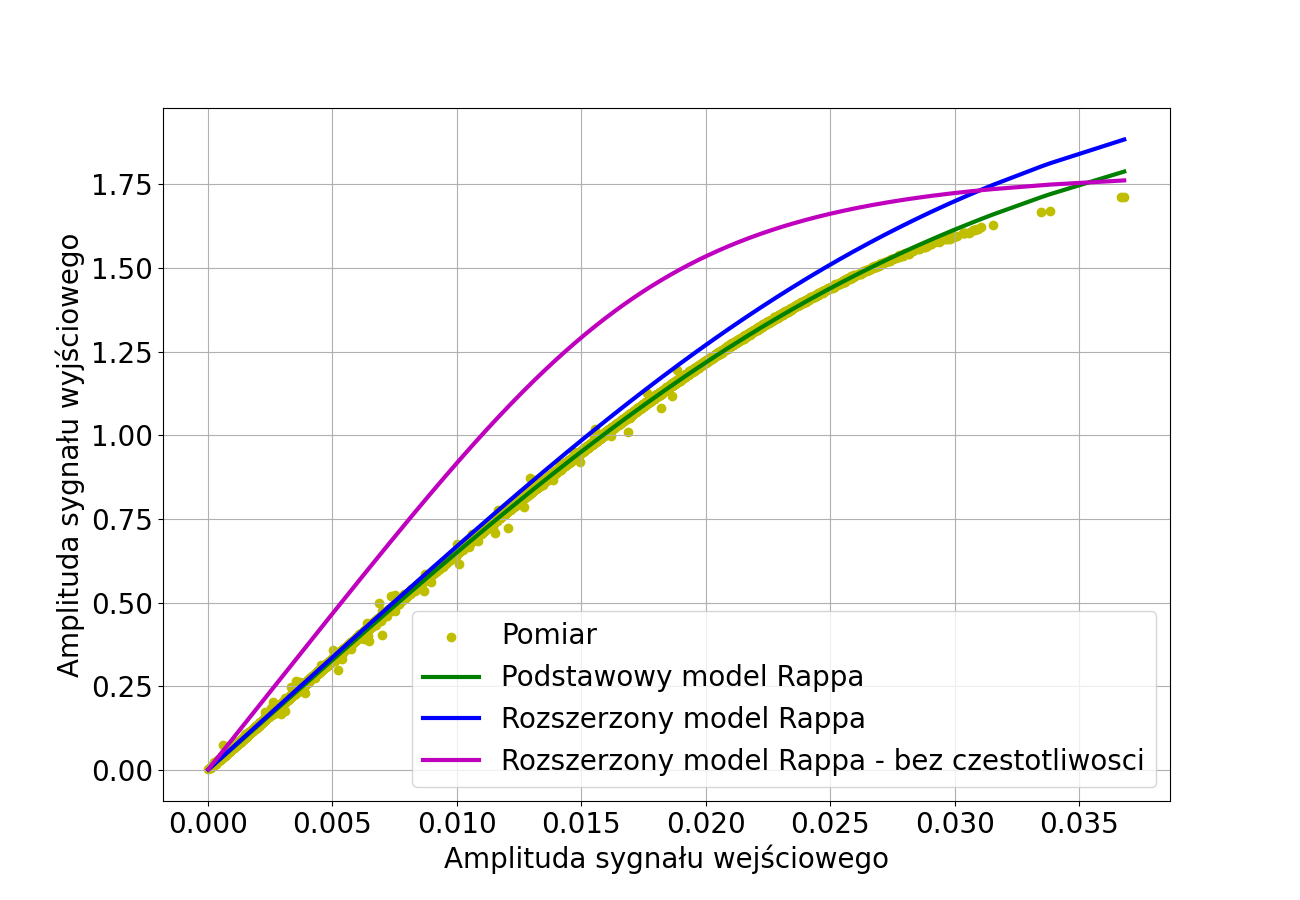}
    \caption{Częstotliwość $f = 2.5GHz$}
    \label{fig:model_59163}
\end{subfigure}
    
\end{minipage}
\caption{Porównanie charakterystyk AM/AM (w V) modeli Rappa dla wzmacniacza ZX60-2534}
\label{fig:models}
\end{figure}

Dla każdego z badanych wzmacniaczy i uzyskanych współczynników $p$, $V_{sat}$, $G$
dopasowano funkcje dwuwymiarowe (\ref{poly_V}), (\ref{eqG}) i (\ref{poly_p2}). Jakość tego dopasowania można ocenić poprzez analizę charakterystyki AM/AM wzmacniacza używając współczynników modelu Rappa: podstawowego (tzn. skalarów pochodzących z dopasowania dla danej częstotliwości i napięcia zasilania), rozszerzonego (tzn. zaproponowanego w tym artykule) oraz rozszerzonego bez częstotliwości (tzn. zaproponowanego w \cite{tafuri_2015_memory_models_for_behavioral_modeling_and_DPD_of_ET_PA}). 
Rysunek \ref{fig:models} przedstawia przykładowe porównanie wykreślone dla wzmacniacza ZX60-2534 dla napięcia zasilania $V_{sup} = 3,6 V$ oraz różnych częstotliwości. Tu warto zaznaczyć, ze model przedstawiony w \cite{tafuri_2015_memory_models_for_behavioral_modeling_and_DPD_of_ET_PA} nie uwzględnia zmiennej częstotliwości, współczynniki dla tego modelu w przypadku wzmacniacza ZX60-2534 zostały wyznaczone dla częstotliwości $f=1.5GHz$. Dla pozostałych wzmacniaczy również wybrano częstotliwości środkowe do wyznaczenia współczynników.

Z analizy wykresu wynika, że zaproponowany rozszerzony model Rappa aproksymuje charakterystyki wzmacniacza w ich liniowym zakresie pracy praktycznie tak dokładnie jak model podstawowy. W każdym przypadku obserwuje się bardziej znaczące różnice podczas modelowania próbek o najwyższych amplitudach. Co ważne, dokładność przybliżenia jest dobra niezależnie od częstotliwości nośnej sygnału nadanego. Tej uniwersalności nie zapewnia model zaproponowany w \cite{tafuri_2015_memory_models_for_behavioral_modeling_and_DPD_of_ET_PA}. Dla częstotliwości $1.5 GHz$, dla której zostały wyznaczone współczynniki Rappa, przybliżenie  wykazują prawie zerowy błąd. Jednak w przypadku pozostałych częstotliwości tzn. 0,5 GHz oraz 2,5GHz,  występują bardzo duże odchylenia. 

\begin{figure}[H]
\centering
\begin{minipage}{0.4\textwidth}
\begin{subfigure}{\textwidth}
    \includegraphics[scale=0.25]{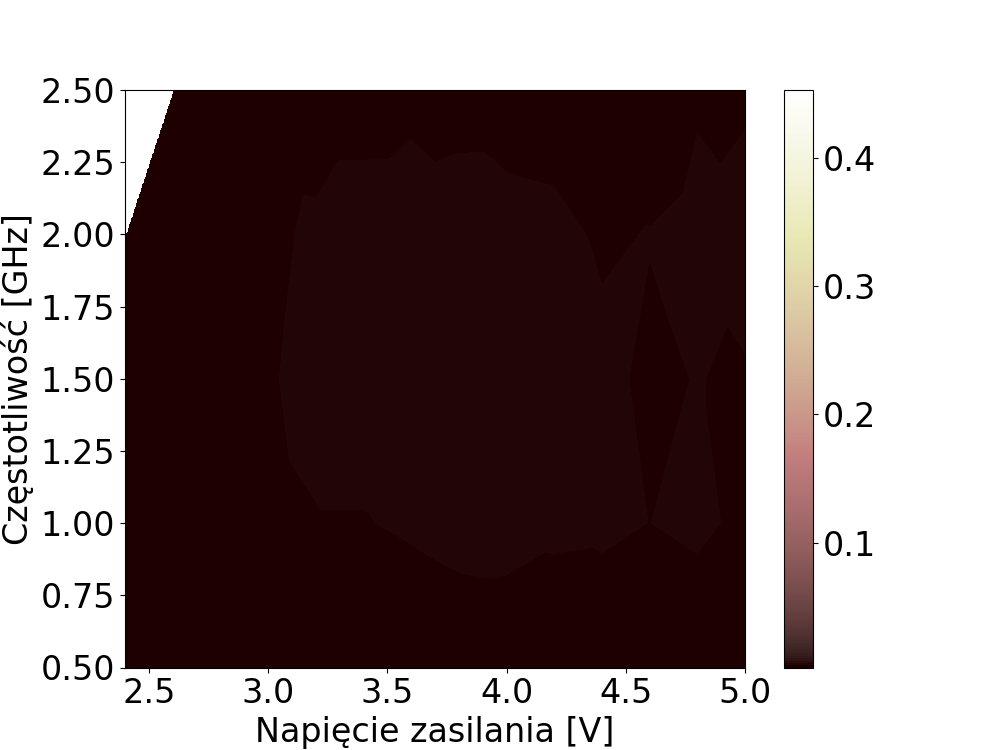}
    \caption{NRMSE dla modelu podstawowego}
    \label{fig:model_20004}
\end{subfigure}
\hfill
\begin{subfigure}{\textwidth}
    \includegraphics[scale=0.25]{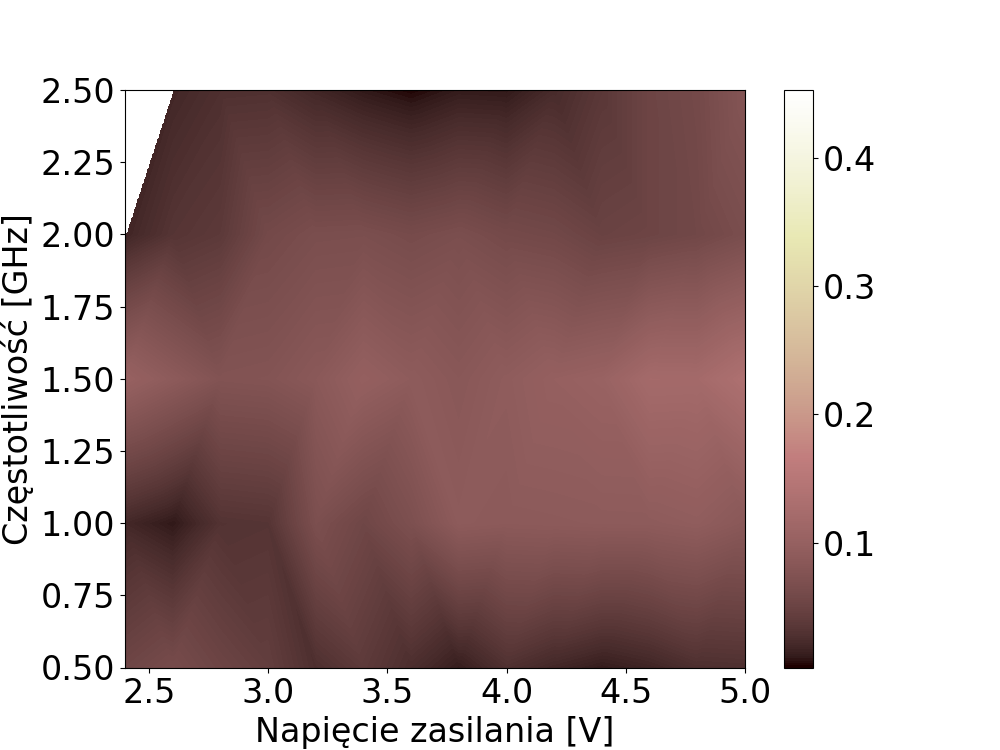}
    \caption{NRMSE dla modelu rozszerzonego}
    \label{fig:model_25345}
\end{subfigure}
\hfill
\begin{subfigure}{\textwidth}
    \includegraphics[scale=0.25]{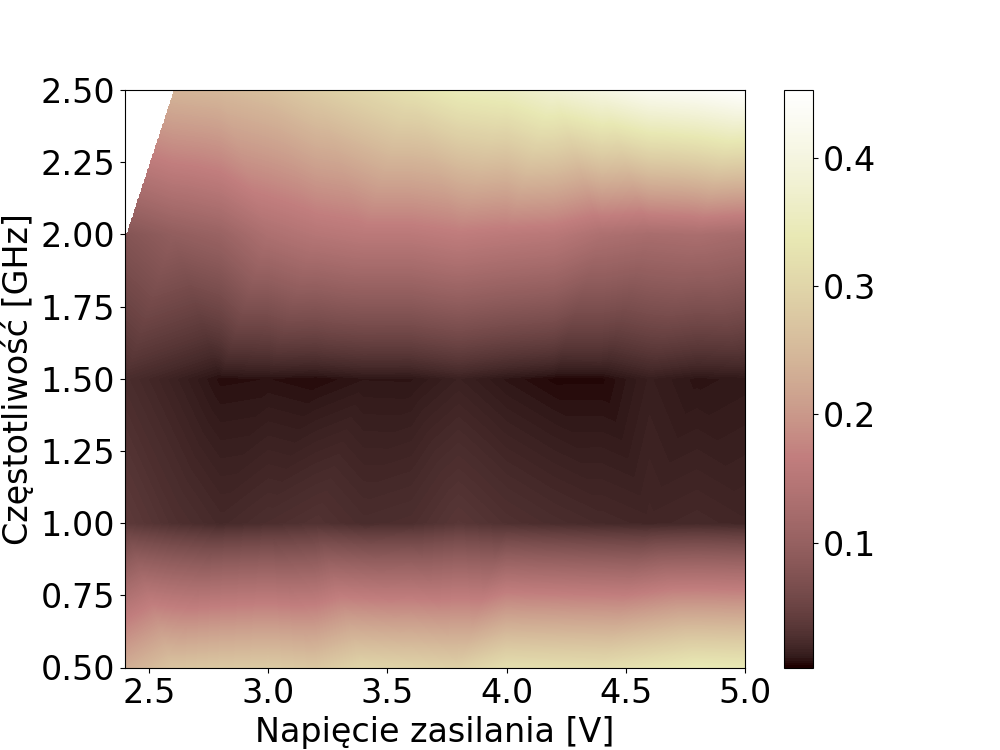}
    \caption{NRMSE dla modelu rozszerzonego bez częstotliwości}
    \label{fig:model_59166}
\end{subfigure}
    
\end{minipage}
\caption{NRMSE dla danych z modelu ZX60-2534}
\label{fig:nmse_2534}
\end{figure}

Pokazuje to znormalizowany (względem wartości zmierzonej charakterystyki AM/AM) pierwiastek ze średniego błędu kwadratowego (ang. \textit{Normalized Root Mean Squared Error}, NRMSE) dopasowania krzywej AM/AM pokazany w funkcji napięcia zasilania i częstotliwości dla wzmacniacza ZX60-2534 na Rys. \ref{fig:nmse_2534} c). Osiągane są wartości do 0.4 (co można interpretować jako $40\%$ błąd w estymowanej wartości wyjściowej charakterystyki AM/AM wzmacniacza). Dla porównania NRMSE dla proponowanego modelu (Rys. \ref{fig:nmse_2534} b))  w całym zakresie działania wzmacniacza rozkłada się równomiernie i nie przekracza $12\%$. Zgodnie z oczekiwaniami, model podstawowy dysponujący niezależnym zestawem parametrów $G$, $V_{sat}$, $p$, dla każdego punktu wykresu daje najdokładniejsze wyniki tzn. NRMSE nie przekracza $0.6\%$.  

Tabela \ref{err_rmse} przedstawiają średnie wartości NRMSE wyliczone na podstawie pełnych danych dla każdego badanego wzmacniacza. Jak widać w przypadku wzmacniacza ZX60-2534 proponowany model wykazuje ponad dwukrotnie mniejsze wartości NRMSE niż model nieuwzględniający częstotliwości. Jednak dla pozostałych wzmacniaczy wartości NRMSE są wyższe. Szczególnie w przypadku modelu ZFL-2000+ błąd otrzymany z zaproponowanego modelu jest znacznie wyższy, co może sugerować inną budowę tego wzmacniacza. 
Trzeba podkreślić, że postaci funkcji aproksymujących współczynniki (np. $p$) wzmacniaczy tj. (\ref{poly_V}), (\ref{eqG}) i (\ref{poly_p2}) były dobierane dla wzmacniacza ZX60-2534. Uzasadnia to bardzo dobre zamodelowanie jego własności przy pomocy niewielkiej liczby współczynników. W przypadku innych wzmacniaczy, choć współczynniki funkcji były dobierane niezależnie struktura tych funkcji była nie zmieniana. Zasadne wydaje się dostosowanie struktury funkcji dla konkretnych urządzeń.


\begin{table}[H]
\centering
\caption{NRMSE dla różnych modeli Rappa }
\label{err_rmse}
\medskip
\begin{tabularx}{\columnwidth}{lX}
\toprule
\textit{} & \textit{Model podstawowy} \\\midrule
    ZFL-2000+  & 0.017825  \\
    ZX60-2534 & 0.003980  \\
    ZX60-5916 & 0.006975  \\  \bottomrule
\textit{} & \textit{Model rozszerzony} \\\midrule
    ZFL-2000+  & 0.476953  \\
    ZX60-2534 & 0.055058  \\
    ZX60-5916 & 0.103444  \\  \bottomrule
\textit{} & \textit{Model rozszerzony bez częstotliwości} \\\midrule
    ZFL-2000+  & 0.118918  \\
    ZX60-2534 & 0.162267  \\
    ZX60-5916 & 0.102723  \\  \bottomrule
\end{tabularx}
\end{table}


\section{Wnioski} \label{sec:conclusions}

W artykule została przedstawiona propozycja rozszerzonego modelu Rappa poprzez wyrażenie jego parametrów jako funkcji wielomianowych parametrów wejściowych: napięcia zasilania wzmacniacza i częstotliwości nośnej przesyłanego sygnału. Wykorzystanie takiego podejścia może otworzyć nowe możliwości modelowania sygnału na wyjściu wzmacniaczy ze zmiennym napięciem zasilania, na przykład dla stacji bazowych używanych w systemach komórkowych następnej generacji. 

Zaproponowany model zapewnia wystarczającą dla wielu celów dokładność opisu charakterystyki nieliniowej przy użyciu niewielkiej liczby współczynników koniecznych do estymacji dla konkretnego wzmacniacza. 


Choć zaproponowany model dobrze oddaje charakterystykę nieliniową dla wzmacniacza dla którego dobierana była funkcja opisująca parametry modelu Rappa, badania wykazały, że jego skuteczność znacząco spada w przypadku innych wzmacniaczy. Wyniki te sugerują, że model nie jest wystarczająco uniwersalny i konieczne jest dostosowanie funkcji opisującej parametry wzmacniacza dla każdego wzmacniacza indywidualnie. Sugerowane są zatem dalsze badania dla znalezienia bardziej uniwersalnego modelu. 


\bibliographystyle{krit}
\bibliography{references}

\begin{thebibliography}{10}
\newcommand{\enquote}[1]{``#1''}

\bibitem{al-kanan_2017_extended_saleh_model_for_behavioral_modeling_ET_PA}
Al-Kanan, Haider, Li, Fu, Tafuri, Felice~Francesco, 2017, \enquote{Extended saleh model for behavioral modeling of envelope tracking power amplifiers}. \emph{2017 IEEE 18th Wireless and Microwave Technology Conference (WAMICON)}, 1--4, IEEE.

\bibitem{al2018hysteresis_nonliearity_model}
Al-kanan, Haider, Tafuri, Felice, Li, Fu, 2018, \enquote{Hysteresis nonlinearity modeling and linearization approach for envelope tracking power amplifiers in wireless systems}. \emph{Microelectronics journal}, 82: 101--107.

\bibitem{boumaiza_2007_systematic_and_adaptive_characterization_for_behavioral_modeling_of_dynamic_nonlinear}
Boumaiza, Slim, Helaoui, Mohamed, Hammi, Oualid, Liu, Taijun, Ghannouchi, Fadhel~M, 2007, \enquote{Systematic and adaptive characterization approach for behavior modeling and correction of dynamic nonlinear transmitters}. \emph{IEEE Transactions on Instrumentation and Measurement}, 56~(6): 2203--2211.

\bibitem{Ghannouchi_behavioral_modeling_and_predistortion_2009}
Ghannouchi, Fadhel~M., Hammi, Oualid, 2009, \enquote{Behavioral modeling and predistortion}. \emph{IEEE Microwave Magazine}, 10~(7): 52--64.

\bibitem{glock2015memoryless}
Glock, Stefan, Rascher, Jochen, Sogl, Bernhard, Ussmueller, Thomas, Mueller, Jan-Erik, Weigel, Robert, 2015, \enquote{A memoryless semi-physical power amplifier behavioral model based on the correlation between am--am and am--pm distortions}. \emph{IEEE Transactions on Microwave Theory and Techniques}, 63~(6): 1826--1835.

\bibitem{Glock2015}
Glock, Stefan, Rascher, Jochen, Sogl, Bernhard, Ussmueller, Thomas, Mueller, Jan-Erik, Weigel, Robert, 2015, \enquote{A memoryless semi-physical power amplifier behavioral model based on the correlation between am–am and am–pm distortions}. \emph{IEEE Transactions on Microwave Theory and Techniques}, 63~(6): 1826--1835.

\bibitem{joung2014survey}
Joung, Jingon, Ho, Chin~Keong, Adachi, Koichi, Sun, Sumei, 2014, \enquote{A survey on power-amplifier-centric techniques for spectrum-and energy-efficient wireless communications}. \emph{IEEE Communications Surveys \& Tutorials}, 17~(1): 315--333.

\bibitem{kim2011optimization_for_ET_shaped_operation_od_ETPA}
Kim, Dongsu, Kang, Daehyun, Choi, Jinsung, Kim, Jooseung, Cho, Yunsung, Kim, Bumman, 2011, \enquote{Optimization for envelope shaped operation of envelope tracking power amplifier}. \emph{IEEE transactions on microwave theory and techniques}, 59~(7): 1787--1795.

\bibitem{kostrzewska2024power}
Kostrzewska, Kornelia, Kryszkiewicz, Pawel, 2024, \enquote{Power amplifier modeling framework for front-end-aware next-generation wireless networks}. \emph{Electronics}, 13~(9): 1643.

\bibitem{kryszkiewicz_2018_amplifier-coupled_tone_reservation}
Kryszkiewicz, Pawel, 2018, \enquote{Amplifier-coupled tone reservation for minimization of ofdm nonlinear distortion}. \emph{IEEE Transactions on Vehicular Technology}, 67~(5): 4316--4324.

\bibitem{kryszkiewicz_2015_obtaining_low_out_band_emission_of_NC-OFDM}
Kryszkiewicz, Pawel, Kliks, Adrian, Bogucka, Hanna, 2015, \enquote{Obtaining low out-of-band emission level of an nc-ofdm waveform in the sdr platform}. \emph{2015 International Symposium on Wireless Communication Systems (ISWCS)}, 66--70, IEEE.

\bibitem{kryszkiewicz_2024_10519547}
Kryszkiewicz, Pawel, Kostrzewska, Kornelia, 2024, \enquote{{Measurements of nonlinearity characteristics and power consumption of 3 power amplifiers (ZFL-2000+,ZX60-2534,ZX60-5916)}}.

\bibitem{li2016new_model_for_ET_PA}
Li, Delong, Yu, Hui, 2016, \enquote{A new model for envelope tracking power amplifier modeling and digital predistortion}. \emph{2016 8th International Conference on Wireless Communications \& Signal Processing (WCSP)}, 1--5, IEEE.

\bibitem{mengozzi2021joint_dual_input_DP}
Mengozzi, Mattia, Angelotti, Alberto~Maria, Gibiino, Gian~Piero, Florian, Corrado, Santarelli, Alberto, 2021, \enquote{Joint dual-input digital predistortion of supply-modulated rf pa by surrogate-based multi-objective optimization}. \emph{IEEE Transactions on Microwave Theory and Techniques}, 70~(1): 35--49.

\bibitem{nota_ZFL-2000+}
Mini-Circuits, ZFL-2000+, \enquote{Power amplifier's datasheet}. \url{"https://www.minicircuits.com/pdfs/ZFL-2000+.pdf"}.

\bibitem{nota_ZX60-2534}
Mini-Circuits, ZX60-2534, \enquote{Power amplifier's datasheet}. \url{"https://www.minicircuits.com/pdfs/ZX60-2534MA+.pdf"}.

\bibitem{nota_ZX60-5916}
Mini-Circuits, ZX60-5916, \enquote{Power amplifier's datasheet}. \url{"https://www.minicircuits.com/pdfs/ZX60-5916MA+.pdf"}.

\bibitem{Nokia_3gpp_Rapp}
Nokia, 2016, \enquote{{Realistic power amplifier model for the New Radio evaluation}}. 3GPP doc. {R4-163314}.

\bibitem{FSL6_manual}
Rohde \&~Schwarz, R\&S~FSL6, \enquote{Manual for the spectrum analyzer}. \url{"https://scdn.rohde-schwarz.com/ur/pws/dl_downloads/dl_common_library/dl_manuals/gb_1/f/sfl_1/FSL_OperatingManual_en_12.pdf"}.

\bibitem{tafuri_2015_memory_models_for_behavioral_modeling_and_DPD_of_ET_PA}
Tafuri, Felice~Francesco, Sira, Daniel, Nielsen, Troels~Studsgaard, Jensen, Ole~Kiel, Mikkelsen, Jan~Hvolgaard, Larsen, Torben, 2015, \enquote{Memory models for behavioral modeling and digital predistortion of envelope tracking power amplifiers}. \emph{Microprocessors and Microsystems}, 39~(8): 879--888.

\bibitem{thota_2020_analysis_of_hybrid_PAPR_reduction_methodth_of_OFDM}
Thota, Sravanti, Kamatham, Yedukondalu, Paidimarry, Chandra~Sekhar, 2020, \enquote{Analysis of hybrid papr reduction methods of ofdm signal for hpa models in wireless communications}. \emph{IEEE Access}, 8: 22780--22791.

\bibitem{thota2020analysis}
Thota, Sravanti, Kamatham, Yedukondalu, Paidimarry, Chandra~Sekhar, 2020, \enquote{Analysis of hybrid papr reduction methods of ofdm signal for hpa models in wireless communications}. \emph{IEEE Access}, 8: 22780--22791.

\bibitem{wang2005design_of_Widebans_ET_PA_for_OFDM_application}
Wang, Feipeng, Yang, Annie~Hueiching, Kimball, Donald~F, Larson, Lawrence~E, Asbeck, Peter~M, 2005, \enquote{Design of wide-bandwidth envelope-tracking power amplifiers for ofdm applications}. \emph{IEEE Transactions on Microwave theory and techniques}, 53~(4): 1244--1255.

\end{thebibliography}

\end{multicols}
\end{document}